\documentclass[journal,onecolumn,12pt]{IEEEtran}

\usepackage{amsmath,amssymb,amsthm,mathtools,graphicx,booktabs,array}
\allowdisplaybreaks
\usepackage[T1]{fontenc}
\usepackage{microtype}
\usepackage[colorlinks=true,linkcolor=blue,urlcolor=blue]{hyperref}
\newtheorem{theorem}{Theorem}
\newtheorem{lemma}[theorem]{Lemma}

\newtheorem{corollary}[theorem]{Corollary}

\newcommand{\tr}{\operatorname{tr}}
\newcommand{\rank}{\operatorname{rank}}
\newcommand{\Vol}{\operatorname{Vol}}
\newcommand{\CN}{\mathcal{CN}}

\newcommand{\CC}{\mathbb C}
\newcommand{\EE}{\mathbb E}
\newcommand{\PP}{\mathbb P}
\newcommand{\norm}[1]{\left\lVert#1\right\rVert}
\title{Universal Decoding for Gaussian MIMO--ISI:\\
Spherical Ensembles and Symmetric Neighborhoods}
\author{Nir Elkayam and Meir Feder\thanks{The authors are with Tel Aviv University, Tel Aviv, Israel.}}
\date{October 22, 2026}
\hypersetup{pdftitle={Universal Decoding for Gaussian MIMO-ISI},pdfauthor={Nir Elkayam and Meir Feder}}
\begin{document}
\maketitle
\begin{abstract}
We construct a universal decoder for cyclic Gaussian multiple-input multiple-output channels with intersymbol interference and unknown channel taps and spatial noise covariance. Codewords are drawn independently and uniformly from a sphere of fixed total energy. Under stated bounds on frequency variation, the decoder attains the ensemble-average error exponent of informed maximum-likelihood decoding for fixed channels and channels with growing memory. We approximate the frequency response by a constant matrix within each frequency block. For this auxiliary channel, every matched log-likelihood is affine in a common sufficient statistic. At fixed received output, we compute the statistic density of an independent competitor. The decoder selects the candidate with the smallest statistic density, equivalently minimizing a weighted sum of residual log-determinants. The density and its support volume give an upper bound on the universal competitor probability. Affinity ensures that at least half of a suitable symmetric neighborhood of the transmitted statistic has matched likelihood no smaller than that of the transmitted word. Integrating a density lower bound over this half gives the matched competitor bound. These bounds establish equality of error exponents, which a Gaussian channel comparison transfers to the original channel. The same decoder attains the matched exponent for independent frequency blocks with prescribed energies satisfying the stated conditions. For energies proportional to block length, an additional gain condition ensures that this exponent is at least that of the total-energy spherical ensemble.
\end{abstract}
\par\noindent\begin{minipage}{\linewidth}
\end{minipage}\par
\section{Introduction}\label{main:sec:introduction}
A receiver with knowledge of a Gaussian channel can decode by choosing
the codeword with the largest likelihood. When the tap matrices and noise
covariance are unknown, we seek a decoder that attains the ensemble-average error exponent of
informed maximum-likelihood (ML) decoding without estimating the channel. We study this question for a cyclic
multiple-input multiple-output (MIMO) channel with finite intersymbol
interference (ISI), using the spherical random coding ensemble in which
each codeword is drawn independently and uniformly with total energy $nP$.
The comparison is with informed ML for this same ensemble. We give
conditions on the channel frequency response that cover fixed channels
and also allow the channel memory to grow with the blocklength.

To understand what a universal decoder must do, consider a transmitted
word and its received output, and draw another codeword independently
from the ensemble. Each decoder has some probability of preferring this
competing word to the transmitted one. Because a random codebook consists
of independent competitors, comparing these probabilities also compares
the ensemble-average decoding errors. The merging approach asks for one decoder whose competitor probability
is at most a subexponential factor times that of each informed decoder.
We state the two bounds needed for this comparison as a good rate-function
property. We then use the channel model to construct a metric that
satisfies these bounds.

Prior work gives two starting points for the MIMO--ISI problem: channels
with memory and decoding metrics for unknown channel parameters.
Ziv~\cite{ziv} constructed a universal decoder for finite-state channels.
Feder and Lapidoth~\cite{feder-lapidoth} gave a general approach for
parametric channel families, including Gaussian ISI.
Lapidoth and Narayan~\cite{lapidoth-narayan} survey coding under channel
uncertainty and the different guarantees that can be obtained.
Here we seek an explicit decoding metric for MIMO--ISI. We also compare
the total-energy sphere with spheres whose energy is fixed in each
frequency block.

Our approach is related to decoding by conditional type classes and
empirical information~\cite{csiszar-korner,merhav1993}, comparison with a
family of decoding metrics~\cite{metricclass}, and merging through
pairwise error probabilities~\cite{elkayam,elkayam-pep}.

Merhav's metric-class framework~\cite{metricclass} constructs universal
decoding metrics from the probabilities of equivalence classes induced by
a family of decoding metrics. Its continuous Gaussian example replaces
exact classes by neighborhoods defined through empirical statistics,
referring to earlier Gaussian analysis~\cite{merhav1993} for their treatment.
That analysis allows a small likelihood loss within an approximate class
and separately controls the resulting error-exponent penalty.
Our proof instead uses the affine dependence of matched log-likelihoods
on a common statistic: reflection within a symmetric feasible neighborhood
guarantees that at least half its volume consists of competitors for each
matched metric. Combined with quantitative density and neighborhood-size
bounds, this gives a direct comparison with matched decoding in the
MIMO--ISI setting.

Lomnitz and Feder~\cite{lomnitz} use determinants of empirical MIMO
matrices to obtain an achievable rate for given input and output sequences.
Their result does not require a probabilistic channel model.
Lapidoth and Telatar~\cite{lapidoth-telatar} establish achievable-rate
optimality of generalized likelihood-ratio decoding for scalar cyclic
Gaussian ISI channels with spherical or independent Gaussian codewords.
Their proof partitions the frequency axis into segments over which
admissible channel responses vary little. We also use a frequency-block
approximation, controlling its error at the error-exponent scale for
a statistic-density decoder in the MIMO setting.
Huleihel and Merhav~\cite{huleihelmerhav} construct a universal decoder
for scalar real Gaussian ISI with unknown taps and noise variance.
It attains the matched random-coding error exponent for a Gaussian coding
distribution restricted to a thin energy shell. Their metric uses a
backward channel in the Fourier domain. Related decoding results for
fading channels appear in~\cite{shayevitz,weinberger}.
Here we study a cyclic complex MIMO channel with unknown
spatial noise covariance and codewords uniform on an exact energy sphere.
At each frequency the channel mixes the transmit antennas through an unknown
matrix, and that matrix can vary across frequencies.
Fix the received output $y$. We seek a \emph{common sufficient statistic}
$u=T_y(x)$ of the candidate $x$ such that every matched log-likelihood
in the family under consideration is affine in $u$. Thus each channel
parameter determines a linear functional of $u$ and an additive constant.
For any feasible pair $u+h,u-h$, the changes in log-likelihood from $u$
are equal in magnitude and opposite in sign.

To construct such a statistic, we transform the channel to the frequency
domain and approximate its response by a constant matrix within each block.
For this block-flat family we obtain an affine representation whose statistic
has dimension proportional to the number of blocks at fixed antenna dimensions.
The statistic records the candidate's input Gram matrix and its projection
onto the received row space in each block. Even if the block energy is
prescribed, it fixes only the trace of the MIMO Gram matrix. The energy
in each transmit direction and the correlations between directions can
still affect the matched likelihood.

We use the density of this statistic to define the universal decoder.
It minimizes a weighted sum of residual log-determinants. The density
also gives an upper bound on its competitor probability. To obtain a
lower bound for each matched decoder, we use the affine log-likelihood
to find competitors near the transmitted statistic. Together, these
bounds give the same error exponent as matched ML in the auxiliary
channel. We then compare the original channel with the block approximation
and show that the approximation does not change either decoder's exponent.
The same decoder is universal for prescribed block energies satisfying
the stated lower-energy condition. When these energies are proportional
to block length, we further show, under the additional gain condition,
that the matched exponent is at least that of the total-energy sphere.

Section~\ref{main:sec:model} states the model and main results.
Section~\ref{main:sec:framework} gives the general comparison tools, and
Section~\ref{main:sec:statistic} constructs the decoder.
Sections~\ref{main:sec:universality} and~\ref{main:sec:energy} prove
universality and compare the ensembles, respectively.

\subsection*{Notation and preliminaries}

For a complex matrix $A$, $A^*$ denotes its conjugate transpose;
$\tr A$, $\det A$, and $\rank A$ denote its trace, determinant,
and rank when defined. We write $\Re z$ for the real part of $z$
and $I$ for an identity matrix of the required size. The Euclidean
vector norm satisfies $\norm v_2^2=v^*v$. For a matrix,
\[
 \norm A_F^2=\sum_{j,k}|A_{jk}|^2=\tr(AA^*),\qquad
 \norm A_{\rm op}^2=\sup_{\norm v_2=1}\norm{Av}_2^2.
\]
Thus the squared Frobenius norm is the sum of the squared entry
magnitudes, while the squared operator norm is the largest possible
squared amplification of a unit vector.

A matrix $A$ is Hermitian if $A=A^*$. For Hermitian matrices,
$A\succeq0$ and $A\succ0$ mean positive semidefinite and positive
definite, respectively: $v^*Av\ge0$ for every $v$, or $v^*Av>0$
for every nonzero $v$. We write $A\preceq B$ when $B-A\succeq0$.
The notation $\lambda_{\min}(A)$ denotes the smallest eigenvalue.
For $A\succeq0$, $A^{1/2}$ denotes its positive semidefinite square root;
$A^{-1/2}$ is used when $A\succ0$.

The Gram matrix $AA^*$ contains the inner products of the rows of
$A$. Its diagonal entries are their squared norms, and
$\tr(AA^*)=\norm A_F^2$. If the columns of $A$ are $a_1,\ldots,a_m$,
the same matrix is $\sum_{j=1}^m a_ja_j^*$, also called the
uncentered scatter matrix. This sum is not divided by $m$.

The law $\CN(0,K)$ on $\CC^d$, for $K\succ0$, is the proper complex
Gaussian law with mean zero, covariance $\EE[ZZ^*]=K$, and density
\[
 p_Z(z)=\frac{\exp(-z^*K^{-1}z)}{\pi^d\det K}.
\]
This density is relative to Lebesgue measure in the real and
imaginary coordinates. In particular, a standard entry
$Z\sim\CN(0,1)$ has independent real and imaginary parts with
variance $1/2$, so $\EE|Z|^2=1$. We use $\PP$ and $\EE$ for
probability and expectation.

Complex matrix spaces are viewed as real Euclidean spaces with
inner product $\Re\tr(A^*B)$. On Hermitian matrices this is
$\tr(AB)$; its induced volume is called Hilbert--Schmidt volume.
For a real affine space $\mathcal A$ with a specified inner product,
$\Vol_{\mathcal A}$ is volume in orthonormal coordinates, and
$\operatorname{relint}\Omega$ is the interior of $\Omega$ within
$\mathcal A$. An unsubscripted norm denotes the norm induced by the
relevant inner product. An affine function is a linear function
plus a constant.

All logarithms are natural. For $b_n>0$, the statements
$a_n=O(b_n)$ and $a_n=o(b_n)$ mean that $|a_n|/b_n$ is eventually
bounded and tends to zero, respectively. For positive sequences,
$a_n\asymp b_n$ means that their ratio is bounded above and away
from zero. We write $a_n\doteq b_n$ when
$n^{-1}\log(a_n/b_n)\to0$. A positive factor is subexponential
when its logarithm is $o(n)$.

\section{Model and Main Results}\label{main:sec:model}
\subsection{Channel and random-coding conventions}
Let $t$ and $s$ be the numbers of transmit and receive antennas.
An input word is a matrix $x\in\CC^{t\times n}$, and the output is a
matrix $y\in\CC^{s\times n}$. Our primary coding prior $Q_1$ is uniform on the total-energy sphere
\begin{equation}\label{main:eq:whole-prior}
 \mathcal S_n=\{x:\norm{x}_F^2=nP\},\qquad P>0.
\end{equation}

For each blocklength, let the memory be $L$, with taps $\mathsf H_0,\ldots,\mathsf H_L\in\CC^{s\times t}$,
and spatial covariance $\Sigma\succ0$. The cyclic finite impulse response
(FIR) channel is
\begin{equation}\label{main:eq:physical}
 Y_\tau=\sum_{\ell=0}^L\mathsf H_\ell X_{\tau-\ell}+N_\tau,
 \qquad N_\tau \sim \mathcal{CN}(0,\Sigma) \quad \text{i.i.d.},
\end{equation}
where indices are modulo $n$. The noise is independent
of the transmitted message and random codebook. The dimensions $t,s$ and power $P$ are fixed. The memory,
taps, and positive-definite covariance may depend on $n$; we suppress
these indices. The receiver
knows the dimensions and the frequency partition defined below, but does
not know the taps or noise covariance.
The energy constraint and rate normalization refer to the $n$ uses of this
cyclic model. A length-$L$ cyclic prefix followed by removal of the prefix
at the receiver produces the same input--output relation over a linear FIR
channel. Its time overhead vanishes for fixed $L$, but its energy must be
accounted for separately: the spherical constraint alone does not make
the energy in the repeated symbols negligible. Extending the result to
linear convolution without a prefix likewise requires control of the
boundary symbols.

We state the random-coding conventions for a general single-codeword
prior $Q$ on length-$n$ codewords. Let
$\mathcal C=\{x_1,\ldots,x_M\}$ be a random codebook with $M=e^{nR}$
words drawn independently from $Q$. The transmitted message is uniform
on $\{1,\ldots,M\}$. Let $g(x,y)$ be a decoding metric. The decoder $D_g$
selects a word maximizing $g(x,y)$, with uniform tie breaking. We write $\overline P_e(g,Q,\theta)$ for its error
probability averaged over the codebook and channel $w_\theta(y\mid x)$.
For matched ML, $g_\theta(x,y)=\log w_\theta(y\mid x)$.

Throughout, we consider $n\to\infty$ at fixed rate and fixed antenna
dimensions and power. The channel parameters may form a deterministic
sequence $\theta=(\theta_n)$; a fixed channel is a special case.
The coding priors, metrics, and decoders form
sequences indexed by $n$; we suppress these indices and integer rounding.
Rates are measured per complex temporal use. We define the error exponent by
\begin{equation}\label{main:eq:exponent}
 \underline E(g,Q,\theta)=\liminf_{n\to\infty}
 -\frac1n\log\overline P_e(g,Q,\theta).
\end{equation}
We use $-\log0=+\infty$, so the exponent takes values in
$[0,+\infty]$. We use a metric and its decoder interchangeably in this notation.

\subsection{Frequency partition and coding ensembles}
Use the frequencies $\omega_k=2\pi k/n$, and partition the unitary discrete Fourier transform (DFT)
coordinates into $B=B_n$ consecutive blocks. Let $I_i$ be the set of frequency indices in block $i$ and $m_i=|I_i|$ its length.
Write $X_i,Y_i$ for the corresponding input and output matrices. The
DFT preserves energy, so $Q_1$ remains uniform on the total-energy sphere
in these coordinates. More blocks improve the frequency approximation.
With fewer blocks the statistic has fewer coordinates, so the volume
of a small neighborhood incurs a smaller penalty in the lower bound on
the matched competitor probability. We use balanced blocks with
\begin{equation}\label{main:eq:scaling}
 \sum_i m_i=n,\quad m_i\asymp n/B,\quad B\to\infty,
 \quad B^2\log n=o(n).
\end{equation}
For example, $B=\lfloor n^{1/3}\rfloor$ with block lengths differing
by at most one satisfies this scaling. The condition on $C_{\theta_n}$ below controls the channel approximation;
$B^2\log n=o(n)$ controls the loss from the neighborhood volume.
For prescribed energies, draw the blocks
independently and uniformly on their spheres:
\begin{equation}\label{main:eq:block-prior}
 \norm{X_i}_F^2=E_i,\qquad \sum_i E_i=nP,\qquad E_i\ge1.
\end{equation}
The lower bound on block energy is assumed for all sufficiently large
$n$. Write $Q_2(\mathbf E)$ for this prior. Each input Gram matrix $X_iX_i^*$ may still vary subject to this trace
constraint.
We first establish universality under this prior and then compare it
with $Q_1$ at the natural energies $E_i=Pm_i$.

\subsection{Channel bounds and main results}
Let $\mathsf H(\omega)=\sum_{\ell=0}^L\mathsf H_\ell
e^{-\mathrm i\ell\omega}$ be the frequency response. Choose a
representative frequency in each block and a numerical constant~$c$
such that every frequency in that block is within $\sqrt{c}/B$ of its
representative, using unwrapped frequency intervals. Balanced blocks
allow the same constant $c$ at every blocklength. Define
\begin{equation}\label{main:eq:channel-constants}
 C_\theta=cP\sup_\omega
 \norm{\Sigma^{-1/2}\mathsf H'(\omega)}_{\rm op}^2,
 \qquad
 G_\theta=\sup_\omega\norm{\Sigma^{-1/2}\mathsf H(\omega)}_{\rm op}.
\end{equation}
With $B$ blocks, $C_\theta/B^2$ bounds the squared error per channel use
in the whitened output mean. The additional constant $G_\theta$ enters
the comparison between coding ensembles: it bounds the change in the
whitened output mean per unit change in the input. For a
channel sequence, these bounds are denoted by $C_{\theta_n}$ and
$G_{\theta_n}$. We require, in addition to~\eqref{main:eq:scaling},
\begin{equation}\label{main:eq:channel-scaling}
 \frac{C_{\theta_n}}{B_n^2}\longrightarrow0.
\end{equation}
For received frequency blocks $Y_i$ of full row rank, define the
candidate residual matrices
\[
 R_i(x,y)=X_i\bigl[I-Y_i^*(Y_iY_i^*)^{-1}Y_i\bigr]X_i^*,
\]
where $X_i$ are the frequency blocks of the candidate $x$.
The decoder chooses
\[
 D_J(y)=\mathop{\arg\min}_{x\in\mathcal C}
 \sum_{i=1}^B(m_i-s-t)\log\det R_i(x,y),
\]
with uniform tie breaking and $\log\det R_i=-\infty$ for singular
residuals. The weights are positive for all sufficiently large $n$
under~\eqref{main:eq:scaling}. On rank-deficient received blocks the
decoder chooses a message uniformly. Positive-definite Gaussian noise
makes this exceptional event null for all sufficiently large $n$.
The receiver needs the partition and antenna dimensions.
Section~\ref{main:sec:statistic} derives this rule from the statistic
density, and Section~\ref{main:sec:decoder} gives its implementation details.

\begin{samepage}
\begin{theorem}[Whole-sphere universality]\label{main:thm:whole}
Let $Q_1$ be the uniform coding prior on the total-energy
sphere~\eqref{main:eq:whole-prior}, and let $D_J$ be the decoder
defined above (see also~\eqref{main:eq:score}). Let $\theta=(\theta_n)$ be a sequence of cyclic MIMO--ISI
channels~\eqref{main:eq:physical} with the stated channel assumptions.
Let the frequency partition and channel bounds satisfy
\eqref{main:eq:scaling} and~\eqref{main:eq:channel-scaling}.
Then
\begin{equation}\label{main:eq:whole-equality}
 \underline E(D_J,Q_1,\theta)=\underline E(\mathrm{ML},Q_1,\theta).
\end{equation}
\end{theorem}
\noindent\emph{Proof:} See Section~\ref{main:sec:universality}.
\end{samepage}

\begin{samepage}
\begin{theorem}[Prescribed-energy universality]
\label{main:thm:block-energy}
Let a channel sequence and frequency partitions satisfy the
assumptions of Theorem~\ref{main:thm:whole}. Let $\mathbf E=(E_1,\ldots,E_B)$
be a prescribed energy profile satisfying~\eqref{main:eq:block-prior},
and let $Q_2(\mathbf E)$ draw the blocks independently and uniformly
on their spheres. Then the decoder $D_J$ satisfies
\begin{equation}\label{main:eq:block-equality}
 \underline E(D_J,Q_2(\mathbf E),\theta)
 =\underline E(\mathrm{ML},Q_2(\mathbf E),\theta).
\end{equation}
\end{theorem}
\noindent\emph{Proof:} See Section~\ref{main:sec:universality}.
\end{samepage}

\begin{samepage}
\begin{theorem}[Natural block-energy nondegradation]
\label{main:thm:nondegradation}
Let a channel sequence and frequency partitions satisfy the
assumptions of Theorem~\ref{main:thm:whole}. Let $Q_1$ be the total-energy
spherical prior and let $Q_2$ draw the blocks independently and uniformly
on spheres with energies $E_i=Pm_i$. Suppose also that
\begin{equation}\label{main:eq:gain-scaling}
 P G_{\theta_n}^2\frac{B_n}{n}\longrightarrow0.
\end{equation}
Then
\begin{equation}\label{main:eq:chain}
 \begin{split}
 \underline E(D_J,Q_2,\theta)&=\underline E(\mathrm{ML},Q_2,\theta)\\
 &\ge\underline E(\mathrm{ML},Q_1,\theta)=\underline E(D_J,Q_1,\theta).
 \end{split}
\end{equation}
\end{theorem}
\noindent\emph{Proof:} See Section~\ref{main:sec:energy}.
\end{samepage}

For a fixed finite-memory channel, $C_\theta$ and $G_\theta$ are finite
constants, so~\eqref{main:eq:scaling} implies both channel conditions.
The results compare ensemble-average error exponents under the same
channel sequence. The third theorem gives a one-sided comparison between
the two ensembles at natural energies; it does not assert strict improvement.

\subsection{Proof roadmap}
Section~\ref{main:sec:framework} develops the probability comparison tools,
and Section~\ref{main:sec:statistic} derives the decoder.
Section~\ref{main:sec:universality} proves Theorems~\ref{main:thm:whole}
and~\ref{main:thm:block-energy} together. Its two approximation costs are
the squared whitened channel error per use, $C_{\theta_n}/B_n^2$,
and the logarithmic neighborhood-volume penalty, $O(B_n^2\log n)$.
Section~\ref{main:sec:energy} proves Theorem~\ref{main:thm:nondegradation}
by rescaling each block to its prescribed energy and controlling the
resulting change in the channel output mean.

\section{General Tools for Universal Decoding}\label{main:sec:framework}
The tools in this section apply to codewords drawn independently from a
common prior and a family of decoding metrics. The density and affine-likelihood hypotheses below
are explicit; their verification for our channel begins in
Section~\ref{main:sec:statistic}.

\subsection{PEP comparison and the good rate-function criterion}
Fix a transmitted word $x$ and output $y$. The conditional pairwise
error probability (PEP) is the probability that an independent
competitor $X'\sim Q$ has at least as large a metric:
\begin{equation}\label{main:eq:weak-pep}
 p_g(x,y)=Q\{g(X',y)\ge g(x,y)\}.
\end{equation}
For the matched metric $g_\theta$, we write $p_\theta$.
For independently drawn codewords, the random-coding union (RCU)
expression bounds the average error probability from above and below,
up to constant factors~\cite[Sec.~III, Lemma~1]{elkayam}. Thus
\begin{equation}\label{main:eq:rcu}
 \overline P_e(g)\doteq
 \EE_{X,Y}\!\left[\min\{1,e^{nR}p_g(X,Y)\}\right],
\end{equation}
where $(X,Y)$ follows the transmitted input--output law. The tie
convention changes only constant factors; Appendix~\ref{app:pairwise}
gives the detailed bounds.

Thus $D_u$ has an error exponent at least as large as that of $D_g$ if
the following bound holds uniformly in $x,y$:
\begin{equation}\label{main:eq:domination}
 p_u(x,y)\le e^{o(n)}p_g(x,y).
\end{equation}
This is the domination criterion of~\cite[Sec.~III, Lemma~2]{elkayam}.
Substitution in~\eqref{main:eq:rcu} gives
$\overline P_e(u)\le e^{o(n)}\overline P_e(g)$. A universal decoder
therefore seeks this comparison with every matched metric in the family.

A useful score $J$ should control the probability of a defeating
competitor through its value. We call the following two bounds the good rate-function property.
The $o(n)$ terms are deterministic and uniform over the input--output
pairs where the bounds are used:
\begin{align}
 p_J(x,y)&\le e^{-J(x,y)+o(n)},\label{main:eq:good-upper}\\
 p_\theta(x,y)&\ge e^{-J(x,y)-o(n)}.
 \label{main:eq:good-lower}
\end{align}
The first bounds the universal decoder's PEP, while the second relates
the same score to every informed decoder. Together they give
\eqref{main:eq:domination}. Since ML is optimal for the given channel,
the resulting exponent comparison is an equality.

The upper bound follows from Markov's inequality whenever
\begin{equation}\label{main:eq:score-normalization}
 \sup_y\EE_{X'}e^{J(X',y)}\le e^{o(n)}.
\end{equation}
This is the moment condition behind the rate-function construction
in~\cite[Appendix~A]{elkayam}. Like a Kraft inequality, it limits the total weight that the score
assigns under the coding prior. We also need the matched lower bound
to compare the universal decoder with the channel family.

The lower bound need only hold below a proof cutoff, provided the
global upper bound makes the remaining error negligible. The next lemma
states this certificate with its finite-blocklength constants. Write
$M_n$ for the message count and choose a fixed finite $R_0$ such that
$\log M_n\le nR_0$; set $c_0=1-e^{-1}$. The precise rounding and
zero-rate convention are given in Appendix~\ref{app:pairwise}.

\begin{samepage}
\begin{lemma}[PEP-to-code-error comparison]
\label{main:prop:criterion}
For codewords drawn independently from a common prior, uniform ties, and
the message counts specified above, suppose $v_n,b_n\ge0$ are $o(n)$,
$\tau_n/n\to\infty$, and measurable $J_n$ satisfy, outside a set null under
the true transmitted joint law,
\begin{align}
 p_J(x,y)&\le\min\{1,e^{-J_n(x,y)+v_n}\},
 \label{main:eq:criterion-upper}\\
 p_\theta(x,y)&\ge e^{-J_n(x,y)-b_n}
 \quad\text{when }J_n(x,y)\le\tau_n.
 \label{main:eq:criterion-lower}
\end{align}
The sequences are deterministic; the inequalities are pointwise in $x,y$.
Then
\begin{align}\label{main:eq:criterion-error}
 \overline P_{e,n}(D_J)
 &\le \frac{2}{c_0}e^{v_n+b_n}\overline P_{e,n}(\mathrm{ML})
       +e^{-\tau_n+nR_0+v_n},
\end{align}
and $\underline E(D_J)=\underline E(\mathrm{ML})$.
\end{lemma}

\end{samepage}

\noindent\emph{Proof idea.} Conditional on the transmitted word, output and tie priority, competing
words are independent. Their union error is comparable to the smaller
of one and the number of competitors times the PEP. Apply the two PEP
bounds below the cutoff; above it, the universal error is
superexponentially small. Appendix~\ref{app:pairwise} gives the
finite-blocklength bounds, tie calculation and exponent argument.

In particular, with $\tau_n=n\log n$ at fixed rate, the conclusion is
\begin{equation}\label{main:eq:retained-error}
 \overline P_e(D_J)
 \le e^{o(n)}\overline P_e(\mathrm{ML})
       +e^{-n\log n+nR+o(n)}.
\end{equation}

The cutoff does not alter the implemented score. It separates the
region where the two PEPs can be compared from a region where the
universal decoder's conditional error is already superexponentially small.

The connection with merging is as follows. Elkayam and
Feder~\cite{elkayam} introduced canonical scores $-\log p_\theta$ and
their merged score $J_{\rm merge}=\sup_\theta[-\log p_\theta]$.
On the retained set the matched lower bound gives
$J_{\rm merge}\le J+b_n$. Normalization controls the competitor
probability of this substitute score. The required inverse-PEP moment bound does not follow from continuity
of the coding prior. Continuity also does not guarantee that the metric
has no atoms. The arbitrary-prior refinement
in~\cite[Sec.~IV-A, Lemma~14]{elkayam-pep} instead bounds a capped
inverse-PEP moment. Appendix~\ref{app:merging} gives the detailed
merging formulas and distinguishes that clipped score from the
unmodified density score used here.

\subsection{The density-score upper bound}
The next bound uses the density and volume with respect to the same
measure. It requires no channel or likelihood structure.

\begin{samepage}
\begin{lemma}[Universal PEP upper bound]\label{lem:universal-pep}
Let $V$ have a normalized, nonnegative finite density~$p$ on a measurable
subset~$\Omega$ of a finite-dimensional real affine space~$\mathcal{A}$,
relative to its induced Euclidean volume. Let $v_n$ be a deterministic
bound on $\log\Vol_{\mathcal A}(\Omega)$, and define $J(u)=-\log p(u)$,
with $J(u)=+\infty$ when $p(u)=0$ and $e^{-\infty}=0$.
Then every $u\in\Omega$ satisfies
\begin{equation}\label{eq:upper}
 \log\Vol(\Omega)\le v_n,\qquad
 p_J(u)\le\min\{1,\exp[-J(u)+v_n]\}.
\end{equation}

Here $p_J(u)=\PP\{J(V)\ge J(u)\}$. At every zero-density point
$p_J(u)=0$, and
$\EE e^{J(V)}=\Vol_{\mathcal A}\{v\in\Omega:p(v)>0\}
\le\Vol_{\mathcal A}(\Omega)$, with equality if $p>0$ almost
everywhere on $\Omega$.
\end{lemma}
\end{samepage}
\begin{proof}
For finite $J(u)$, a winning competitor has $p(v)\le p(u)$, so
\[
 p_J(u)=\int_{\{v\in\Omega:J(v)\ge J(u)\}}p(v)\,dv
 \le e^{-J(u)}\Vol(\Omega).
\]

Combining this with the volume bound and $p_J\le1$ proves the claim.
Since a zero-density set has probability zero, $\PP\{J(V)=+\infty\}=0$.
Only a competitor in that set can tie or defeat an infinite transmitted
score, proving $p_J(u)=0$ there. On $p>0$, multiplying $e^J=1/p$ by
the density gives one; integration proves the moment assertion.
\end{proof}

\subsection{The symmetric-neighborhood lower bound}
For the lower bound, affinity of the matched metric and three local
conditions suffice. All volumes and densities again use the same measure.

\begin{samepage}
\begin{lemma}[Matched PEP from a symmetric neighborhood]
\label{main:lem:neighborhood}
Let $V$ have density function $p(v)$ supported on $\Omega \subseteq \mathcal{A}$, where
$\mathcal{A}$ is a finite-dimensional real affine space of positive dimension.
Fix $u \in \Omega$ with $p(u) > 0$, and let $\mathcal{N}$ be a full-dimensional
ball in $\mathcal{A}$ centered at $u$. Let $b \ge 0$, $c \ge 0$, and suppose
\begin{enumerate}
\item $\mathcal N\subseteq\Omega$;
\item $p(v)\ge e^{-c}p(u)$ for $v\in\mathcal N$;
\item $\Vol_{\mathcal A}(\mathcal N)\ge e^{-b}$.
\end{enumerate}
Then for every affine function $\ell$ on $\mathcal A$,
\begin{equation}\label{main:eq:generic-neighborhood}
 \PP\{\ell(V)\ge\ell(u)\}\ge\tfrac12 e^{-b-c}p(u).
\end{equation}
\end{lemma}
\end{samepage}

\begin{proof}
Affinity gives $\ell(u+h)-\ell(u)=-[\ell(u-h)-\ell(u)]$.
If $\ell$ is nonconstant, the reflection $u+h\mapsto u-h$ therefore
exchanges the two half-balls and preserves their volume. Their common boundary has zero volume, so each has half
the ball's volume. Integrate the density lower bound over the winning
half. If $\ell$ is constant, integrate over the whole ball instead.
\end{proof}

\begin{figure}[!htb]
\centering
\includegraphics[width=0.62\columnwidth]{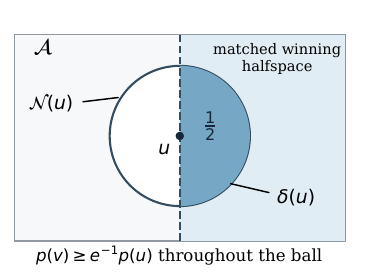}
\caption{A symmetric neighborhood in the space defined by the energy constraint.
If the matched log-likelihood is nonconstant on this space, its comparison
with the transmitted word divides the neighborhood into two equal parts.
One part has likelihood at least as large as that of the transmitted
statistic $u$. A lower bound on the density there gives a lower bound
on the matched PEP.}
\label{main:fig:halfball}
\end{figure}

The nonconstant affine functional bisects volume. The density lower
bound then converts the winning half-volume into probability; no
symmetry of the density is assumed. Figure~\ref{main:fig:halfball}
illustrates this step.

\subsection{Gaussian channel comparison}
We use Gaussian comparison to replace the physical channel by its block
approximation and, later, to adjust codeword energies. Within each
comparison the noise covariance is common to both channels. Holding a
decoder fixed compares its errors; applying the comparison to each
channel's ML rule also compares their optimal errors.

\begin{samepage}
\begin{lemma}[Gaussian comparison]\label{lem:gaussian-transfer}\label{main:lem:gaussian}
Let $\mathcal C=\{x_1,\ldots,x_M\}$ be a given finite codebook,
and let the transmitted message be uniform on $\{1,\ldots,M\}$.
Let $f_1,f_2$ be given functions and let $\mathsf K\succ0$ be a covariance
matrix. Define two channels by applying $f_j$ to the transmitted word
and adding Gaussian noise: when message $m$ is transmitted, channel $j$
produces $Y_j=f_j(x_m)+W$, where $W\sim\CN(0,\mathsf K)$ and $j=1,2$.
The message indices distinguish codewords even if their vectors coincide.
Suppose
\[
 \max_m\norm{\mathsf K^{-1/2}[f_1(x_m)-f_2(x_m)]}_2\le\Delta,
\]
and let $D$ be a fixed measurable decoder, using the same independent tie
randomness on both channels, and let $e_j$ be its error probability on
channel $j$. Then, for every $\nu>1$,
\begin{equation}\label{eq:gaussian-holder}
 e_1\le e^{(\nu-1)\Delta^2} e_2^{1-1/\nu}.
\end{equation}
The same bound holds for the respective optimal ML errors and for
ensemble-average errors under a common codebook distribution when the
displacement bound is uniform over its support.
\end{lemma}

\end{samepage}

\noindent\emph{Proof idea.} Whiten the noise and compare the two Gaussian likelihoods. Their
likelihood-ratio moment is bounded by the squared displacement of their
means. H\"older's inequality compares the error probabilities for a fixed
rule; applying it to each experiment's ML rule also compares the optimal
errors. Appendix~\ref{app:transfer} gives the calculation.

\begin{corollary}[Gaussian exponent comparison]\label{cor:gaussian-exponent}
Suppose two error sequences satisfy, for every $\nu>1$,
\begin{equation}\label{eq:gaussian-exponent-condition}
 e_{1,n}\le e^{(\nu-1)\Delta_n^2}
             (K_ne_{2,n})^{1-1/\nu},
\end{equation}
where $\Delta_n^2=o(n)$, $K_n\ge1$, and $\log K_n=o(n)$.
For $E_j=\liminf_n[-n^{-1}\log e_{j,n}]$, with $-\log0=+\infty$,
\begin{equation}\label{eq:gaussian-exponent}
 E_1\ge E_2.
\end{equation}
\end{corollary}
\begin{proof}
Fix $\nu>1$. Taking negative logarithms, dividing by $n$, and taking
lower limits gives $E_1\ge(1-1/\nu)E_2$. Let $\nu\to\infty$.
This also covers $E_2=+\infty$.
\end{proof}

\section{Construction of the Universal Decoder}\label{main:sec:statistic}
\subsection{The auxiliary block channel}\label{main:sec:reduction}
Apply the unitary discrete Fourier transform (DFT),
$\widehat X_k=n^{-1/2}\sum_{\tau=0}^{n-1}X_\tau e^{-2\pi\mathrm i k\tau/n}$,
and likewise transform $Y$ and $N$. For $k=0,\ldots,n-1$, cyclic
convolution becomes exactly
\begin{equation}\label{main:eq:dft}
 \widehat Y_k=\mathsf H(\omega_k)\widehat X_k+\widehat N_k,
 \quad\mathsf H(\omega)=\sum_{\ell=0}^L\mathsf H_\ell e^{-\mathrm i\ell\omega},
 \quad\omega_k=2\pi k/n.
\end{equation}
It preserves total input energy and the noise law: transformed columns
remain independent with covariance $\Sigma$. In particular, the spherical
prior remains the same spherical prior in Fourier coordinates.

The DFT channel still has a different gain at every frequency. Using the partition specified in Section~\ref{main:sec:model},
replace it
within block $i$ by $H_i=\mathsf H(\bar\omega_i)$ at a representative
frequency $\bar\omega_i$ in that block.
This defines the auxiliary block channel
\begin{equation}\label{main:eq:block}
 Y_i=H_iX_i+N_i.
\end{equation}
Unlike the DFT itself, this replacement is an approximation.

Section~\ref{main:sec:approximation} bounds its approximation error.
For now, we use the exact block model to derive the decoder.

\subsection{From ML to the sufficient statistic}
Conditional on any transmitted word, positive-definite Gaussian noise
gives each received block a density on $\CC^{s\times m_i}$. Since
$m_i\ge s$ eventually, all received blocks have full row rank almost
surely under both the physical and auxiliary channels. Throughout the
construction we work on this event. On exceptional outputs, the decoder
chooses a message uniformly.

Fix the received blocks $Y_i$ in the exact auxiliary channel. For a
candidate with blocks $X_i$, the Gaussian log-likelihood is
$-\sum_i\tr[\Sigma^{-1}(Y_i-H_iX_i)(Y_i-H_iX_i)^*]$ plus a
candidate-independent constant. Expanding the square gives
\begin{equation}\label{main:eq:ml-expansion}
 \ell_\theta(x;y)=c_\theta(y)
 -\sum_i\tr(H_i^*\Sigma^{-1}H_iX_iX_i^*)
 +2\Re\sum_i\tr(H_i^*\Sigma^{-1}Y_iX_i^*).
\end{equation}
Thus the candidate enters through its Gram matrices $X_iX_i^*$ and
its correlations $X_iY_i^*$ with the output. We express those
correlations in orthonormal coordinates to simplify the density calculation.

\begin{samepage}
\begin{lemma}[Affine sufficient statistic]\label{main:lem:statistic}
Let $Y_i$ be fixed received blocks of full row rank $s$, and let
$W_i\in\CC^{m_i\times s}$ have orthonormal columns spanning their row
spaces. For a candidate $x$ with realized frequency blocks $x_i$, define
\begin{equation}\label{main:eq:statistic}
 f_i=x_ix_i^*,\qquad z_i=x_iW_i.
\end{equation}
The common sufficient statistic is $T_y(x)=((f_i,z_i))_{i=1}^B$.
For a random codeword $X$ with blocks $X_i$, its statistic is
$T_y(X)=((F_i,Z_i))_{i=1}^B$, where $F_i=X_iX_i^*$ and $Z_i=X_iW_i$.
Every matched auxiliary log-likelihood is affine in the realized statistic:
\begin{align}\label{main:eq:affine}
 \ell_\theta(f,z;y)&=c_\theta(y)
 -\sum_i\tr(H_i^*\Sigma^{-1}H_i f_i)
 +2\Re\sum_i\tr(D_i^*z_i),\\[-2pt]
 D_i&=H_i^*\Sigma^{-1}Y_iW_i.\nonumber
\end{align}
\end{lemma}
\end{samepage}
\begin{proof}
Since $Y_i=Y_iW_iW_i^*$, we have
$x_iY_i^*=z_i(Y_iW_i)^*$. Substituting in
\eqref{main:eq:ml-expansion} gives~\eqref{main:eq:affine}.
\end{proof}

$f_i$ records input energy and spatial correlations; $z_i$ records the
candidate's projection onto the received row space. No orthogonal
residual is needed to evaluate the matched metric from $(f,z)$.

To compute the statistic density, decompose each candidate into its
projection and orthogonal component:
\[
 x_i=z_iW_i^*+x_{i,\perp},\qquad
 x_{i,\perp}=x_i(I-W_iW_i^*).
\]
Orthogonality gives $f_i=z_iz_i^*+x_{i,\perp}x_{i,\perp}^*$. Hence
\begin{equation}\label{main:eq:derived-residual}
 R_i=R_i(f_i,z_i):=f_i-z_iz_i^*=x_{i,\perp}x_{i,\perp}^*.
\end{equation}
This is determined by the sufficient statistic. It measures the candidate
component orthogonal to the output row space within the frequency block.
On the full-rank event it can be computed as
\begin{equation}\label{main:eq:residual}
 R_i=x_i[I-Y_i^*(Y_iY_i^*)^{-1}Y_i]x_i^*.
\end{equation}
For scalar input and output, with block energy $e_x$,
\[
 R_i=e_x(1-\eta_i),\qquad
 \eta_i=\frac{|x_iY_i^*|^2}{e_x\norm{Y_i}_F^2}.
\]
Stronger alignment with the output reduces the residual at fixed block
energy. On the total-energy sphere, the block energies can vary while
their sum remains fixed. We will see that the statistic density has a
simple expression in terms of these residuals.

For a scalar block with nonzero output, the projection is the least-squares
backward fit $x_i=\widehat\beta_iY_i+x_{i,\perp}$, where
$\widehat\beta_i=x_iY_i^*/\norm{Y_i}_F^2$. For matrix inputs,
$\tr R_i$ measures total residual energy, while $\det R_i$ combines
residual spread across spatial directions.

\subsection{Domain and reference volume}
Put $E=nP$, $k_i=m_i-s$, and $a_i=k_i-t>0$ eventually.
We first fix the space on which density and volume will be computed.

\begin{samepage}
\begin{lemma}[Statistic domain and reference volume]\label{lem:statistic-domain}
Fix full-row-rank outputs $Y_i$ and row-space bases $W_i$, with $m_i-s\ge t$.
Let $f_i$ range over Hermitian $t\times t$ matrices and
$z_i\in\CC^{t\times s}$. Let $\mathcal A$ be their affine space
subject to either $\sum_i\tr f_i=E$ or $\tr f_i=E_i$ for every
block, with $E_i>0$ and $\sum_i E_i=E>0$.
For either energy constraint, the feasible statistic set is exactly
$\Omega=\{(f,z)\in\mathcal A:f_i-z_iz_i^*\succeq0\ \forall i\}$.
It has nonempty relative interior in $\mathcal A$. The whole-sphere
dimension is $B(t^2+2ts)-1$; the block-sphere dimension is
$B(t^2+2ts-1)$. Density and neighborhood volumes use the
same Euclidean measure induced on $\mathcal A$ by
$\sum_i\bigl[\tr(f_ig_i)+\Re\tr(z_i^*w_i)\bigr]$.
\end{lemma}
\end{samepage}

\begin{proof}
For any fixed full-row-rank output, choose
$W_i\in\CC^{m_i\times s}$ with orthonormal columns spanning its row
space. Complete it to
a unitary matrix $[W_i\ U_i]$. Multiplication by this unitary matrix
preserves rectangular real Euclidean volume. It also gives
$x_i=[x_iW_i\ x_iU_i][W_i\ U_i]^*$ and
$R_i=(x_iU_i)(x_iU_i)^*$. This proves residual positivity and identifies
the part of the input whose scatter it measures.

Conversely, for any $(f,z)$ satisfying the selected trace constraints and
$f_i-z_iz_i^*\succeq0$, factor this residual as $v_iv_i^*$ with
$v_i\in\CC^{t\times(m_i-s)}$. Such a factor exists because
$m_i-s\ge t$: a square root can be padded with zero columns.
Then $x_i=[z_i\ v_i][W_i\ U_i]^*$ realizes exactly $f_i,z_i$ and
the required energy. This realizes every point of the asserted support.

On the ambient space use
$\langle(f,z),(g,w)\rangle=\sum_i\bigl[\tr(f_ig_i)+\Re\tr(z_i^*w_i)\bigr]$.
A Hermitian $t$ by $t$ matrix has real dimension $t^2$, and $z_i$ has
dimension $2ts$. One independent total-trace equation removes one
dimension; independent block-trace equations remove $B$. This gives
the two stated affine dimensions. Positive scalar Grams with the
required traces and $z=0$ show that the relative interior is nonempty.
All volumes below are induced by this one inner product.

\end{proof}

Write the two spaces explicitly as
\begin{samepage}
\begin{align}
 \mathcal A_{\mathrm w}&=\{(f,z):\sum_i\tr f_i=E\},\nonumber\\
 \Omega_{\mathrm w}&=\{(f,z)\in\mathcal A_{\mathrm w}:R_i\succeq0\},\nonumber\\
 d_{\mathrm w}&=B(t^2+2ts)-1.\label{eq:whole-support}
\end{align}
\end{samepage}

\begin{align}
 \mathcal A_{\mathrm b}&=\{(f,z):\tr f_i=E_i\ \text{for every }i\},\nonumber\\
 \Omega_{\mathrm b}&=\{(f,z)\in\mathcal A_{\mathrm b}:R_i\succeq0\},\nonumber\\
 d_{\mathrm b}&=B(t^2+2ts-1).\label{eq:block-support}
\end{align}

Both dimensions are positive. Basis invariance and boundary details
are given in Appendix~\ref{app:geometry}.

\subsection{Statistic densities under both ensembles}\label{main:sec:densities}
We now compute the density of $T_y(X')$ at fixed output $y$, with $X'$
drawn independently from the coding prior. We use the reference measure
just defined on the selected trace space. This is the competitor density,
not the posterior density of the transmitted input given $y$. First use
standard Gaussian blocks, then normalize their energy globally or block
by block.
Let $V\in\CC^{t\times k}$ have independent $\CN(0,1)$ entries.
The scatter matrix $Q=VV^*$ has the complex Wishart law: it is
the sum of the outer products of $k$ independent Gaussian columns
with covariance $I_t$. Thus $\EE Q=kI_t$; no division by $k$ is
included. For $k\ge t$, $Q$ is positive definite almost surely,
and its standard density~\cite{goodman} is proportional to
\[
 e^{-\tr q}\det(q)^{k-t},\qquad q\succ0.
\]
Changing the fixed reference volume changes only its normalizing
constant. The independent Gaussian projection and orthogonal component
give the following joint density. We use normalized densities throughout,
but do not need to evaluate their constants explicitly.

\begin{samepage}
\begin{lemma}[Single-block Gaussian statistic density]
\label{lem:gaussian-statistic}
Let $G\in\CC^{t\times m}$ have independent $\CN(0,1)$ entries, and
let $W\in\CC^{m\times s}$ be a fixed matrix with orthonormal columns.
Suppose $k=m-s\ge t$. Define
\begin{equation}\label{eq:prime}
 F'=GG^*,\qquad Z'=GW,\qquad R'=F'-Z'Z'^*.
\end{equation}
Relative to Hilbert--Schmidt volume for $f$ and real Euclidean volume
for $z$, the joint density is
\begin{equation}\label{eq:single-gaussian}
 p_{F',Z'}(f,z)=C_{m,s,t}e^{-\tr f}
       \det(f-zz^*)^{m-s-t},\quad f-zz^*\succ0.
\end{equation}
Here $C_{m,s,t}>0$ depends only on the dimensions.
\end{lemma}

\end{samepage}

\noindent\emph{Proof idea.} Rotate the Gaussian input into its projection onto the received row
space and its orthogonal component. These components are independent.
Their Gaussian and Wishart densities multiply; replacing the residual
by the full Gram matrix preserves volume. Appendix~\ref{app:gaussian-density}
gives the calculation.

For independent blocks, their densities multiply. Normalizing their
energy globally or independently in each block gives both spherical
densities.

\begin{samepage}
\begin{lemma}[Spherical statistic density]\label{lem:spherical-density}
Let $G_i$ be independent standard Gaussian blocks as in
Lemma~\ref{lem:gaussian-statistic}, let $S_i=\norm{G_i}_F^2$, and put
$S=\sum_iS_i$. Global normalization $X_i=\sqrt{E/S}\,G_i$ gives the
whole-sphere law, whose statistic density on $\Omega_{\rm w}$ is
\begin{equation}\label{eq:whole-density}
 p_{\mathrm w}(f,z)=C_{\mathrm w}\prod_i\det R_i^{a_i}.
\end{equation}

Independent normalization $X_i=\sqrt{E_i/S_i}\,G_i$ gives the
block-sphere law for every positive energy vector with $\sum_iE_i=E$.
Its statistic density on $\Omega_{\rm b}$ is
\begin{equation}\label{eq:block-density}
 p_{\mathrm b}(f,z)=C_{\mathrm b}\prod_i\det R_i^{a_i}.
\end{equation}

The positive constants $C_{\rm w}$ and $C_{\rm b}$ normalize the
respective densities. They depend on the dimensions and prescribed
energies, not on output magnitudes. Both densities use their own induced
affine volume from Lemma~\ref{lem:statistic-domain} and are zero on
singular residuals.
\end{lemma}
\end{samepage}

\noindent\emph{Proof idea.} Gaussian direction is independent of energy.
Normalizing the input retains only its direction. In the joint density
of energy and normalized statistic, the exponential and all coordinate
scaling factors depend only on energy. Integrating out energy therefore
changes only the density constant. Apply the same argument separately
to each block for block normalization. Appendix~\ref{app:spherical-density}
gives the change of coordinates for every prescribed positive energy.

The two priors have the same determinant powers but different trace
constraints and density constants. Section~\ref{main:sec:universality}
bounds each constant using normalization on a small interior ball.
The decoder needs only the density ordering; it does not evaluate the
normalizing constant or the proof cutoff.

Each density is positive in the relative interior. A singular residual
is the zero set of a nonzero determinant polynomial on the affine space,
so the boundary has zero volume and zero competitor probability for
every fixed output. Appendix~\ref{app:geometry} gives the details.

\subsection{Definition and implementation of the score}\label{main:sec:decoder}
Define the normalized proof scores $J_{\rm w}=-\log p_{{\rm w},y}$ and
$J_{\rm b}=-\log p_{{\rm b},y}$, evaluated at $T_y(x)$. For a selected
ensemble write $p,C$ for its density and constant and $J$ for its score:
\begin{equation}\label{eq:J}
 J(u)=-\log p(u)=-\log C-\sum_i a_i\log\det R_i(u),
\end{equation}

The block contributions in this logarithmic score add. Under independent
block spheres, $p_{\rm b}(u)=\prod_i p_{{\rm b},i}(u_i)$, so
$J_{\rm b}=\sum_i[-\log p_{{\rm b},i}(u_i)]$. Each contribution already contains
the dimension weight $a_i$; no further block-length scaling is needed.
For the whole-energy sphere, block energies are coupled. Its score still
has the additive determinant form in~\eqref{eq:J}, by the exact density
formula, rather than by independence of the blocks.

Set $J=+\infty$ at singular residuals. The common implemented decoder is
\begin{equation}\label{main:eq:score}
 D_J(y)=\mathop{\arg\min}_{x\in\mathcal C}
       \sum_i a_i\log\det R_i(x,y),
\end{equation}
with uniform ties. It maximizes the selected normalized score because
the normalizer is candidate-independent. The receiver needs the
partition, antenna dimensions, and received row spaces, but no channel
gains or noise covariance. Singular competitors have probability zero
at fixed output, as required for the infinite-score case of
Lemma~\ref{lem:universal-pep}.

An invertible spatial transformation of the output leaves the score
unchanged.
For any invertible $s\times s$ matrix $A$, replacing each $Y_i$ by $AY_i$
preserves its row space and hence every candidate's residual $R_i$ and
decoding score. In particular, whitening by $A=\Sigma^{-1/2}$ leaves the
decoder's decision unchanged while replacing the taps by
$\Sigma^{-1/2}\mathsf H_\ell$ and the noise covariance by the identity.
The receiver therefore does not need to perform or estimate this whitening
transformation.
The probability bounds and proof cutoff use a normalized density, whose
constant is bounded in Section~\ref{main:sec:universality}. The decoder
only compares candidates and does not evaluate this constant.

In particular, probability bounds for the block-sphere ensemble use
$J_{\rm b}$ and its block-trace volume, even though the implemented
decisions agree with those obtained from $J_{\rm w}$.
For scalar input at fixed block energy, greater alignment with the output
reduces the residual and increases the score. For matrix inputs, the
determinant combines the residual eigenvalues across spatial directions.

\section{Proof of the Universality Theorems}\label{main:sec:universality}
We prove Theorems~\ref{main:thm:whole} and~\ref{main:thm:block-energy}
together. We bound competitor probabilities using the normalized scores
$J_{\rm w}$ and $J_{\rm b}$ on their respective affine trace spaces.
Their candidate-dependent terms are identical. The receiver therefore
uses~\eqref{main:eq:score} for either ensemble without evaluating a normalizing constant. For the universal PEP bound, we need an upper
bound on the support volume. For the matched PEP bound, we need a ball
inside the support with lower bounds on its density and volume. The
resulting PEP losses are $O(B\log n)$ and $O(B^2\log n)$, respectively.
We first bound the channel approximation, then verify these two PEP bounds.

Select either ensemble and use its own density $p$, normalizer $C$,
affine space $\mathcal A$ and support $\Omega$ from
Section~\ref{main:sec:statistic}. Thus $J$ means $J_{\rm w}$ or
$J_{\rm b}$ consistently, including in the cutoff. Put $E=nP$.
Every density and volume uses the same induced measure on the selected
affine space. We use the full-row-rank event established in
Section~\ref{main:sec:statistic}.

\subsection{Approximation of the physical channel}
\label{main:sec:approximation}\label{sec:reduction}
\begin{samepage}
\begin{lemma}[Block approximation]\label{lem:block-approximation}
Let $H_i=\mathsf H(\bar\omega_i)$ at the prescribed representative
frequency in each block, with maximum frequency distance
$w_n\le\sqrt{c}/B_n$. For any coding support of total energy $nP$,
\begin{align}
 \Delta_n^2&:=\sup_x\sum_i\sum_{k\in I_i}
 \norm{\Sigma^{-1/2}[\mathsf H(\omega_k)-H_i]\widehat X_k}_2^2
 \nonumber\\
 &\le nC_{\theta_n}/B_n^2.\label{eq:isi-displacement}
\end{align}
Thus $C_{\theta_n}/B_n^2\to0$ gives $\Delta_n=o(\sqrt n)$ for both
coding ensembles.
\end{lemma}
\end{samepage}

\begin{proof}
Integrating the whitened derivative on the unwrapped interval gives
\[
 \norm{\Sigma^{-1/2}[\mathsf H(\omega_k)-H_i]}_{\rm op}
 \le w_n\sup_\omega\norm{\Sigma^{-1/2}\mathsf H'(\omega)}_{\rm op}.
\]
Square this bound, multiply by $\norm{\widehat X_k}_2^2$, and sum.
The unitary DFT gives $\sum_k\norm{\widehat X_k}_2^2=nP$.
For reference, the elementary tap bound
\[
 \sup_\omega\norm{\mathsf H'(\omega)}_{\mathrm{op}}
 \le C_{\rm der}:=\sum_{\ell=0}^L\ell\norm{\mathsf H_\ell}_{\mathrm{op}}
\]
also gives $\Delta_n^2\le
\norm{\Sigma^{-1/2}}_{\rm op}^2C_{\rm der}^2w_n^2nP$.
All these inequalities hold separately at every blocklength.
\end{proof}

For either ensemble,
\begin{equation}\label{main:eq:sequence-transfer}
 \frac{\Delta_n^2}{n}\le\frac{C_{\theta_n}}{B_n^2}\longrightarrow0.
\end{equation}
Apply Corollary~\ref{cor:gaussian-exponent} in both
directions with $K_n=1$. Lemma~\ref{lem:gaussian-transfer} supplies its
probability bound for any fixed decoder and for the respective ML
decoders. Their physical and auxiliary liminf exponents therefore agree.

\subsection{The universal PEP upper bound}\label{sec:pep}
At fixed output, write $u=T_y(x)$ and let $V=T_y(X')$ be an independent
competitor statistic. Both the density score and the auxiliary matched
metric depend on the candidate through this statistic.

\begin{samepage}
\begin{lemma}[Support-volume upper bound]\label{lem:support-volume}
For either ensemble there is a deterministic $v_n=O(B\log n)$,
independent of the output and channel, such that
$\log\Vol_{\mathcal A}(\Omega)\le v_n$.
\end{lemma}
\end{samepage}

\begin{proof}
Positivity gives $f_i\succeq0$ and $z_iz_i^*\preceq f_i$. Thus
$\norm{f_i}_F\le\tr f_i$ and $\norm{z_i}_F^2\le\tr f_i$, whence
\[
 \sum_i\norm{f_i}_F^2\le E^2,\qquad
 \sum_i\norm{z_i}_F^2\le E.
\]
The support lies in an ambient ball of radius $\sqrt{E^2+E}=O(n)$.
Its section by $\mathcal A$ is a $d$-dimensional ball of no greater
radius. In orthonormal coordinates a containing cube has volume at most
$(2\sqrt{E^2+E})^d$. Since $d=O(B)$, its logarithm is $O(B\log n)$.

\end{proof}

Lemma~\ref{lem:universal-pep} now gives
\begin{equation}\label{main:eq:universal-integral}
 p_J(u)\le p_y(u)\Vol(\Omega).
\end{equation}

\begin{equation}\label{main:eq:support-bound}
 \log\Vol(\Omega)\le O(B\log n)=o(n),\qquad
 p_J(u)\le e^{-J(u)+o(n)}.
\end{equation}

The positive interior density and null boundary also give
$\EE_{X'}e^{J(X',y)}=\Vol_{\mathcal A}(\Omega)$, verifying
the moment condition. Thus the first good rate-function bound holds
globally, with the same deterministic $v_n=O(B\log n)$ for every
full-row-rank output.

\subsection{Verification of the neighborhood conditions}
First construct a feasible ball with stable density at any interior
point. We use it twice: at a fixed reference point to bound the density
constant, and below the score cutoff to obtain the matched PEP bound.
All constants depend only on the fixed dimensions, power and block
balancing constants, uniformly over the admissible energies and outputs.

\begin{samepage}
\begin{lemma}[Neighborhood feasibility]\label{lem:ball}
For $u \in \operatorname{relint}(\Omega)$, put
\begin{equation}\label{eq:radius}
 \rho(u)=\min_i\lambda_{\min}(R_i(u)),\qquad
 \delta(u)=\frac{\min\{1,\rho(u)\}}{16tn(1+\sqrt E)}.
\end{equation}
The full-dimensional ball in the energy affine space
\begin{equation}\label{eq:neighborhood}
 \mathcal{N}(u) = \bigl\{ u + h : h \in \mathcal{A} - u,\; \|h\| \le \delta(u) \bigr\}
\end{equation}
is contained in $\operatorname{relint}(\Omega)$.
\end{lemma}

\end{samepage}

\begin{proof}
Every perturbation is tangent to $\mathcal A$, so it preserves the
required trace or traces. For $v=u+(\Delta f,\Delta z)$,
\[
 R_i(v)-R_i(u)=\Delta f_i-z_i\Delta z_i^*
 -\Delta z_i z_i^*-\Delta z_i\Delta z_i^*.
\]
Using $\norm{z_i}_{\mathrm{op}}\le\sqrt E$ and $\delta\le1$ gives
\begin{align*}
 \norm{R_i(v)-R_i(u)}_{\mathrm{op}}
 &\le(1+2\sqrt E)\delta+\delta^2\\
 &\le3(1+\sqrt E)\delta\le\rho(u)/(4tn).
\end{align*}
Set $\chi=1/(4tn)$. Since $R_i(u)\succeq\rho(u)I$, we obtain
\begin{equation}\label{eq:relative}
 (1-\chi)R_i(u)\preceq R_i(v)
 \preceq(1+\chi)R_i(u).
\end{equation}
All residuals are positive definite. Lemma~\ref{lem:statistic-domain}
therefore makes every point of this entire affine ball feasible.
\end{proof}

\begin{samepage}
\begin{lemma}[Local density stability]\label{lem:density-stability}
For the ball of Lemma~\ref{lem:ball}, $p(v)\ge e^{-1}p(u)$ throughout
$\mathcal N(u)$, uniformly over the output and auxiliary channel.
\end{lemma}

\end{samepage}

\begin{proof}
Multiply~\eqref{eq:relative} on both sides by $R_i(u)^{-1/2}$.
All eigenvalues of the resulting relative matrix lie in
$[1-\chi,1+\chi]$, so the log-determinant ratio lies between
$t\log(1-\chi)$ and $t\log(1+\chi)$. The common density constant cancels.
Put $a_{\rm tot}=\sum_i a_i$. Since $a_i>0$, $a_{\rm tot}\le n$,
and $\chi\le1/4$,
\[
 \log\frac{p(v)}{p(u)}\ge ta_{\rm tot}\log(1-\chi)
 \ge-2tn\chi=-\frac12\ge-1.
\]
The upper bound is also at most $ta_{\rm tot}\log(1+\chi)\le1/4$.
Thus feasibility and density stability use the same radius.
\end{proof}

\noindent\emph{A bound on the density constant.}
Write $p(u)=C\prod_i\det R_i(u)^{a_i}$. Choose the reference point $u_0=(f^0,z^0)$ with
$z_i^0=0$, $f_i^0=(e_i^0/t)I_t$, with $e_i^0=Pm_i$ for the whole
sphere and $e_i^0=E_i$ for block spheres. Eventually $e_i^0\ge1$,
so this point is in the relative interior and its residual eigenvalues
are at least $1/t$. The radius in~\eqref{eq:radius} is therefore of
order $n^{-3/2}$ at $u_0$. For $d=\dim\mathcal A=O(B)$, an inscribed
cube in this ball gives
$\log\Vol(\mathcal N(u_0))\ge-O(B\log n)$.
Normalization and local density stability imply
\[
 1\ge\int_{\mathcal N(u_0)}p(v)\,dv
 \ge e^{-1}p(u_0)\Vol(\mathcal N(u_0)).
\]
Also, $\sum_i a_i\log\det R_i(u_0)\ge-tn\log t$. Thus
\begin{equation}\label{eq:normalizer-bound}
 \log C\le O(B\log n)+tn\log t=O(n).
\end{equation}
This uses $E_i\ge1$ for block spheres and the existing block scaling.
It requires no score cutoff or residual eigenvalue floor.

Now retain the region
\begin{equation}\label{eq:cutoff}
 \mathcal G_n=\{u\in\Omega:J(u)\le n\log n\}.
\end{equation}
Finite $J$ implies that every residual is positive definite, so every
point of $\mathcal G_n$ is in the relative interior. The next estimate
prevents the ball radius from becoming too small on this region.

\begin{samepage}
\begin{lemma}[Residual eigenvalue lower bound]\label{lem:residual-floor}
For either ensemble, if $J(u)\le n\log n$, every residual is positive
definite and
$\rho(u):=\min_i\lambda_{\min}R_i(u)\ge
\exp[-C_\rho B\log n]$ for all sufficiently large $n$.
The constant is uniform over the output and auxiliary channel.
\end{lemma}

\end{samepage}

\noindent\emph{Proof idea.} Total energy bounds all residual eigenvalues
from above. Isolate the smallest one in the weighted determinant
product and use the score cutoff together with~\eqref{eq:normalizer-bound}.
Appendix~\ref{app:residual-floor} gives the eigenvalue calculation.

\begin{samepage}
\begin{lemma}[Neighborhood-volume lower bound]\label{lem:neighborhood-volume}
For $u\in\mathcal G_n$, the ball of Lemma~\ref{lem:ball} has
$\Vol_{\mathcal A}(\mathcal N(u))\ge\exp[-O(B^2\log n)]$, uniformly.
\end{lemma}
\end{samepage}

\begin{proof}
Let $d=\dim\mathcal A$, and let $\kappa_d$ be the volume of the unit
ball in $\mathbb R^d$. Orthonormal coordinates on $\mathcal A$ give
$\Vol(\mathcal N(u))=\kappa_d\delta(u)^d$.
There is no loss from intersection with the feasible support, since
the whole ball lies in it. Lemma~\ref{lem:residual-floor} gives
$\log(1/\delta(u))=O(B\log n)$ uniformly. The cube
$[-1/\sqrt d,1/\sqrt d]^d$ lies in the unit ball, so
\[
 \kappa_d\ge\left(\frac{2}{\sqrt d}\right)^d,\qquad
 \log\kappa_d\ge d\log2-\frac d2\log d
 \ge-O(d\log(d+1)).
\]
Using $d=O(B)$ proves
\[
 \log\Vol(\mathcal N(u))
 =\log\kappa_d+d\log\delta(u)\ge-O(B^2\log n).\qedhere
\]
\end{proof}

The preceding lemmas establish the three conditions of
Lemma~\ref{main:lem:neighborhood}. In particular, they retain the bounds
\begin{equation}\label{eq:local}
 \Vol(\mathcal N(u))=\kappa_d\delta(u)^d,\qquad
 p(v)\ge e^{-1}p(u)\quad(v\in\mathcal N(u)).
\end{equation}
Apply Lemma~\ref{main:lem:neighborhood} to the affine matched likelihood
of Lemma~\ref{main:lem:statistic}. For $u\in\mathcal G_n$,
\begin{align}\label{main:eq:matched-bound}
 p_\theta(u)&\ge\frac1{2e}e^{-J(u)}\kappa_d\delta(u)^d\nonumber\\
 &\ge e^{-J(u)-\beta_n},\qquad
 \beta_n=O(B^2\log n)=o(n).
\end{align}
A constant likelihood uses the whole ball and satisfies the same bound.
The loss is uniform over the selected ensemble's admissible energies
and full-row-rank outputs. Its constants depend only on the fixed dimensions,
power and block balancing constants, not on output magnitudes or
matched likelihood coefficients.

For clarity, for the block-sphere ensemble these two bounds read
\begin{align}
 p_{J_{\rm b}}(x,y)&\le e^{-J_{\rm b}(x,y)+O(B\log n)},
 \label{main:eq:block-upper}\\
 p_\theta(x,y)&\ge e^{-J_{\rm b}(x,y)-O(B^2\log n)}
 \quad\text{if }J_{\rm b}(x,y)\le n\log n.
 \label{main:eq:block-lower}
\end{align}

Here both bounds use $J_{\rm b}$ and volume in $\mathcal A_{\rm b}$;
the ball preserves every block trace. The whole-sphere proof uses
$J_{\rm w}$ and $\mathcal A_{\rm w}$. This verifies the good
rate-function property for each ensemble with its own normalized score.

\subsection{Completion of the universality proof}
\label{main:sec:completion}\label{app:completion}
The two bounds hold with deterministic losses $v_n=O(B\log n)$ and
$b_n=O(B^2\log n)$, both $o(n)$, uniformly in the auxiliary channel
sequence. Apply Lemma~\ref{main:prop:criterion} with
$\tau_n=n\log n$: the matched bound is used on $J\le n\log n$,
and the global universal bound controls $J>n\log n$, including
$J=+\infty$.
It gives the error comparison~\eqref{main:eq:retained-error}, including
the superexponential contribution above the cutoff. ML optimality then
gives equality of the auxiliary universal and matched liminf exponents.

Use the same implemented residual-determinant decoder as a map of the
received vector and labelled codebook in the two channels. Apply
Corollary~\ref{cor:gaussian-exponent} in both directions to this fixed
rule, using~\eqref{main:eq:sequence-transfer} for the displacement. This gives
\[
 \underline E(D_J;\mathrm{physical})
 =\underline E(D_J;\mathrm{auxiliary})
 =\underline E(\mathrm{ML};\mathrm{auxiliary})
 =\underline E(\mathrm{ML};\mathrm{physical}).
\]

This proves both theorems for liminf exponents in $[0,+\infty]$.
The block prior changes the affine space and
normalizer; the geometric estimates hold in that space, and the physical
comparison uses only the common total energy.

For an exact block channel, no frequency approximation is needed. The
same proof covers fixed $B$ and separate positive-definite covariances
$\Sigma_i$ in each block: replacing $\Sigma$ by $\Sigma_i$ in the
quadratic expansion preserves affinity. The long-block condition controls
the local volume loss, while $C_{\theta_n}/B_n^2\to0$ is used only
to compare the auxiliary and physical channels.

\section{Comparison of the Two Ensembles}\label{main:sec:energy}
\label{sec:energy-comparison}
Theorems~\ref{main:thm:whole} and~\ref{main:thm:block-energy} already
give the two same-prior equalities in Theorem~\ref{main:thm:nondegradation}.
It remains to prove $\underline E(\mathrm{ML},Q_2,\theta)\ge
\underline E(\mathrm{ML},Q_1,\theta)$ for the natural profile $E_i=Pm_i$.
We make this comparison on the physical channel, with the same message
count, by conditioning a single-word draw and then changing its block radii.

\subsection{Exact natural-energy coupling}
\begin{samepage}
\begin{lemma}[Exact natural-energy coupling]\label{lem:radial-coupling}
Let $E=nP$ and $X\sim Q_1$. Write
\[
 \Lambda_i=\norm{X_i}_F^2/E,\qquad X_i=\sqrt{E\Lambda_i}\,\Theta_i,
\]
where $\Theta_i$ is the block direction. Put $e_i=m_i/n$ and define
\begin{equation}\label{eq:radial-new}
 \Phi_n(X)_i=\sqrt{Ee_i}\,\Theta_i.
\end{equation}
Then $\Phi_n(X)$ has the natural block-sphere law $Q_2$. The same
is true after conditioning $X$ on any positive-probability event that
depends only on $\Lambda$. Independently mapped codewords remain iid.
\end{lemma}
\end{samepage}

\begin{proof}
Generate $X$ by globally normalizing independent standard Gaussian
blocks $G_i$. Each block has a uniform direction $\Theta_i$, independent
of its squared radius $S_i=\norm{G_i}_F^2$. The blocks are independent,
so all directions are independent of all radii and hence of
$\Lambda_i=S_i/\sum_j S_j$. An energy-only event changes no direction
law. Replacing the radii by $\sqrt{Ee_i}$ therefore gives exactly $Q_2$.
Zero block energies are null; arbitrary directions may be assigned there.
A deterministic map applied separately to independent draws preserves
their independence.
\end{proof}

\subsection{Movement bound}
The exact image law alone gives no error comparison. We next control
how far the coupling moves a word, and hence its conditional output mean.
This is where the additional gain condition is used.

\begin{samepage}
\begin{lemma}[Energy-adjustment displacement]\label{lem:energy-movement}
For the map of Lemma~\ref{lem:radial-coupling}, let
$\varepsilon_n=(B-1)/(tn+1)$ and
$\mathfrak d=\norm{\Phi_n(X)-X}_F^2/E
=\sum_i(\sqrt{\Lambda_i}-\sqrt{e_i})^2$.
The energy fractions satisfy
\begin{align}
 \EE \Lambda_i&=e_i,\qquad
 \EE \Lambda_i^2=\frac{tm_i(tm_i+1)}{tn(tn+1)},\nonumber\\
 \operatorname{Var}(\Lambda_i)&=\frac{e_i(1-e_i)}{tn+1}.
 \label{eq:energy-moments}
\end{align}

Moreover, $\EE[\mathfrak{d}]\le\varepsilon_n$. For $B>1$, the energy-only
event $\mathcal E_n=\{\mathfrak d\le2\varepsilon_n\}$ has
probability at least $1/2$. Let
$\Delta_{n,\mathrm{move}}$ be the largest whitened conditional-mean
displacement between $x$ and $\Phi_n(x)$ on this event. Then
\begin{equation}\label{eq:mean-movement}
 \Delta_{n,\mathrm{move}}^2
 \le 2G_{\theta_n}^2 nP\varepsilon_n.
\end{equation}
For $B=1$ the map is the identity. Under the stated gain scaling the
right side is $o(n)$.
\end{lemma}

\end{samepage}

\noindent\emph{Proof idea.} The Gaussian block radii give the energy-fraction moments. Rationalizing
the square roots bounds the expected squared movement by the weighted
variance sum. Markov's inequality gives an energy-only event with a
uniform displacement bound. Appendix~\ref{app:movement} gives the
moment and displacement calculations.

In particular, $\EE[\mathfrak{d}]\le
\sum_i\operatorname{Var}(\Lambda_i)/e_i=\varepsilon_n$.
Markov's inequality then gives the retained event. On this event,
multiplying the squared input displacement by $G_{\theta_n}^2$ bounds
the squared whitened mean displacement:
\begin{equation}\label{main:eq:movement-gain}
 \frac{\Delta_{n,\rm move}^2}{n}
 \le 2P G_{\theta_n}^2\varepsilon_n\longrightarrow0.
\end{equation}

Indeed, $\varepsilon_n=(B_n-1)/(tn+1)\le B_n/(tn)$, so
\eqref{main:eq:gain-scaling} gives $\Delta_{n,\rm move}^2/n\to0$.
Write
$\pi_n=Q_1(\mathcal E_n)\ge1/2$; if $B=1$, take $\mathcal E_n$ to
be the full support, with $\pi_n=1$.

\subsection{Conditioning and error comparison}
\begin{samepage}
\begin{lemma}[Conditioning a coding prior]\label{lem:conditioning}\label{main:lem:conditioning}
Let $Q$ be a coding prior and $\mathcal E$ a measurable input event with
$Q(\mathcal E)=b>0$. Draw codewords independently from $Q$ or from
$Q(\cdot\mid\mathcal E)$, respectively.
For the same channel, measurable metric, message count, and uniform ties,
\begin{equation}\label{eq:conditioning}
 \overline P_e(D,Q(\cdot\mid\mathcal E))
 \le b^{-2}\overline P_e(D,Q).
\end{equation}
\end{lemma}
\end{samepage}

\begin{proof}
Give each codeword an independent uniform priority, with smaller
priorities winning ties. This induces uniform selection among tied
labels. Conditional on transmitted priority~$\xi$, a tied competitor
wins with probability~$\xi$. At fixed
transmitted $(x, y)$ and priority~$\xi$, write the competitor winning
probability as $p = p_>(x, y) + \xi p_=(x, y)$. Conditioning the competitor on~$\mathcal{E}$ gives $p_{\mathcal{E}} \le p/b$. With $N=M-1$, let
$\Psi_N(v)=1-(1-v)^N$, extended constantly as 1 for $v>1$. It is
nondecreasing and concave with $\Psi_N(0)=0$, so
$\Psi_N(p_{\mathcal E})\le b^{-1}\Psi_N(p)$.
The exact conditional ensemble error is $\Psi_N(p)$. Integrating over the
transmitted input restricted to~$\mathcal{E}$ costs a second factor~$b^{-1}$; integrating output and priority proves \eqref{eq:conditioning}.
The resulting ensemble draws its codewords independently from
$Q(\cdot\mid\mathcal E)$. The two factors account for the transmitted
draw and the independent competing draw.
\end{proof}

\noindent\emph{Completion of Theorem~\ref{main:thm:nondegradation}.}
We first condition the single-word prior $Q_1$ on $\mathcal E_n$, then
apply $\Phi_n$ separately to the independently drawn codewords.
Conditioning gives a uniform bound on how far each word moves. Since
$\mathcal E_n$ depends only on block energies, the directions remain
independent and uniform, so the moved words have exactly the natural
block-sphere law $Q_2$.

Lemma~\ref{lem:energy-movement} gives a uniform whitened displacement
$\Delta_{n,\mathrm{move}}^2=o(n)$. Pair each labelled codebook
$\mathcal C$ drawn from $Q_1(\cdot\mid\mathcal E_n)$ with
$\Phi_n\mathcal C$. Apply Lemma~\ref{lem:gaussian-transfer} with the
original ML label rule held fixed, followed by ML optimality for the
moved codebook. Its ensemble-average conclusion and
Lemma~\ref{lem:conditioning} give, for $\nu>1$,
\begin{equation}\label{eq:energy-final-error}
 \begin{split}
 \overline P_{e,\mathrm{ML}}(Q_2)
 &\le e^{(\nu-1)\Delta_{n,\mathrm{move}}^2}\\
 &\quad\cdot[\pi_n^{-2}\overline P_{e,\mathrm{ML}}(Q_1)]^{1-1/\nu}.
 \end{split}
\end{equation}
Corollary~\ref{cor:gaussian-exponent} applies with
$K_n=\pi_n^{-2}\le4$ and $\Delta_n^2=\Delta_{n,\mathrm{move}}^2=o(n)$.
It gives $\underline E(\mathrm{ML},Q_2,\theta)\ge
\underline E(\mathrm{ML},Q_1,\theta)$, including infinite exponents.
Theorems~\ref{main:thm:whole} and~\ref{main:thm:block-energy} complete \eqref{main:eq:chain} and the proof of
Theorem~\ref{main:thm:nondegradation}. The direction of the comparison comes from
conditioning the original prior and then using optimality after movement.

For other admissible prescribed profiles,
Theorem~\ref{main:thm:block-energy} gives the same-prior universal/ML
equality; the ensemble comparison above concerns the natural profile.

\section{Discussion}\label{main:sec:discussion}
\subsection{Fixed and growing memory}
The channel enters the universality proof through $C_\theta$, and the
ensemble comparison also uses $G_\theta$. For fixed $P>0$,
\eqref{main:eq:gain-scaling} permits $G_{\theta_n}=o(\sqrt{n/B_n})$,
subject to the frequency-variation condition of Theorem~\ref{main:thm:whole}. The statistic has $O(B)$
coordinates regardless of the tap count. No grid over channel parameters
is used: all auxiliary matched log-likelihoods are affine in the same
statistic, so one symmetric-neighborhood bound applies to all of them.
The frequency partition approximates frequency variation.

For a fixed finite-memory channel and $\Sigma\succ0$, both channel
bounds are finite. Thus $B_n\to\infty$ and $B_n^2\log n=o(n)$ suffice
for all three theorems. For growing memory, the tap bounds
\begin{align}\label{main:eq:tap-bounds}
 C_{\theta_n}&\le cP\left(
 \sum_{\ell=0}^{L_n}\ell\norm{\Sigma_n^{-1/2}
 \mathsf H_{\ell,n}}_{\rm op}\right)^2,\nonumber\\
 G_{\theta_n}&\le\sum_{\ell=0}^{L_n}
 \norm{\Sigma_n^{-1/2}\mathsf H_{\ell,n}}_{\rm op}
\end{align}
follow by the triangle inequality applied to the frequency response
and its derivative, and give simple sufficient conditions. If the sum of whitened tap norms
is uniformly bounded, then $C_{\theta_n}=O(L_n^2)$ and
$G_{\theta_n}=O(1)$. Choosing $L_n=o(B_n)$ and
$B_n^2\log n=o(n)$ satisfies both channel conditions.
Such a block count exists whenever $L_n^2\log n=o(n)$.
These are sufficient conditions; the bounds do not establish
the largest possible memory growth.

Since the blocks are balanced, each exponent $a_i=m_i-s-t$ in the
density factor $\det(R_i)^{a_i}$ is of order $n/B$. The logarithm of the
statistic-support volume is bounded above by $O(B\log n)$. The analytical cutoff $\tau_n=n\log n$ makes the error above
the cutoff superexponentially small at fixed rate. These quantities
do not introduce further channel assumptions. The bounds $C_\theta$
and $G_\theta$ control the comparisons, rather than determine the ML
exponent itself.

\subsection{Computational cost}
For bounded antenna dimensions, forming the block statistics and determinants
costs $O(n)$ per candidate. Fourier transformation costs $O(n\log n)$ for
the output and for each candidate unless codewords are stored in Fourier
coordinates. An exhaustive search evaluates the score for all $M$
candidates, multiplying the per-candidate arithmetic by $M$.

\subsection{Codebook guarantees and limitations}
Our guarantees concern ensemble-average error along a given channel
sequence, including a fixed channel. For a finite family $\Theta_n$,
let $\mu_{j,n}$ be the ensemble-average error of $D_J$ at channel $j$.
If $\log|\Theta_n|=o(n)$, there is a deterministic codebook such that
$P_e(\mathcal C,j)\le e^{n\eta_n}\mu_{j,n}$ for every $j\in\Theta_n$,
where $n\eta_n=\log|\Theta_n|+\sqrt n$ and $\eta_n\to0$.
Appendix~\ref{app:finite-family} gives the Markov and union-bound proof.
Extending this conclusion to a continuous family requires an
approximation argument with uniform bounds on the error-exponent scale.

The proof needs long balanced blocks with $B^2\log n=o(n)$ so that
the neighborhood volume does not change the error exponent. Open
questions include whether the result holds with shorter frequency blocks
than the sufficient conditions above allow, whether one codebook can achieve a uniform guarantee
over a continuous channel family, and whether finite-tap generalized likelihood-ratio decoding attains the
matched exponent.

\appendices
\section{General Comparison Tools}
\subsection{PEP-to-code-error comparison}\label{app:pairwise}
We give the finite-blocklength details of the PEP-to-code-error
comparison. Write $Q_n$ for the selected prior and take $M_n\ge2$
with $\log M_n\le nR_0$ for some finite $R_0$. A precise rate-$R$
convention is $M_n=\max\{2,\lceil e^{nR}\rceil\}$; this keeps a
nontrivial code at zero rate. Put $c_0=1-e^{-1}$. The certificate permits different deterministic sublinear losses
in the upper and lower PEP bounds.

\begin{proof}[Proof of Lemma~\ref{main:prop:criterion}]
Fix $n$, the transmitted word $x$, and output $y$. Random codebook
symmetry permits us to call the transmitted label 1: conditional on its
value, the other $N=M_n-1$ words are still independent draws from the
same prior. Giving each label an independent uniform priority implements
uniform selection among all labels tied for the largest metric.
Conditional on the transmitted priority~$\xi$, a competitor wins either
by having a strictly larger metric or by tying and having priority below
$\xi$. Its probability is therefore $p_{g,\xi}=p_{g,>}+\xi p_{g,=}$,
where $p_{g,>}$ and $p_{g,=}$ denote the strict and tied comparisons.
The competing words and their priorities are independent under this
conditioning, so the probability that at least one wins is exactly
$\Psi_N(p_{g,\xi})=1-(1-p_{g,\xi})^N$.

Here are the elementary estimates behind the comparison. The union bound
gives $\Psi_N(p)\le\min(1,Np)$. Conversely $(1-p)^N\le e^{-Np}$,
and $1-e^{-z}\ge(1-e^{-1})\min(1,z)$ for $z\ge0$: on $[0,1]$
this follows from concavity, and on $[1,\infty)$ from monotonicity.
Write $c_0=1-e^{-1}$. The weak probability $p_g=p_{g,>}+p_{g,=}$
satisfies $\xi p_g\le p_{g,\xi}\le p_g$. Furthermore
$\min(1,\xi z)\ge\xi\min(1,z)$ for $0\le\xi\le1$.
Integration over $\xi$ thus proves
\[
 \frac{c_0}{2}\min(1,Np_g)\le
 \int_0^1\Psi_N(p_{g,\xi})\,d\xi\le\min(1,Np_g).
\]
This step avoids an assumption that matched ties have probability zero.

We can enlarge $v_n,b_n$ to nonnegative sequences without changing their
order. Set $K_n=e^{v_n+b_n}\ge1$. On $J\le\tau_n$, the assumed
inequalities give $p_J\le K_np_\theta$. Since
$\min(1,K_nz)\le K_n\min(1,z)$, conditional universal error is at most
$K_n\min(1,Np_\theta)$, and hence at most $2K_n/c_0$ times the
priority-averaged matched conditional error. On $J>\tau_n$ it is at most
$Np_J\le e^{-\tau_n+nR_0+v_n}$. Integrate the first inequality over
the retained set under the true joint law, and enlarge the matched error
integral to the full space. Integrating the second inequality costs no
more than its uniform bound. True-joint null exceptional sets contribute
zero. This proves \eqref{main:eq:criterion-error}.

For completeness, put $q_n=\overline P_{e,n}(\mathrm{ML})$ and let
$r_n$ denote the additive term. The proved inequality is
$\overline P_{e,n}(D_J)\le e^{o(n)}q_n+r_n$, where
$-n^{-1}\log r_n\to\infty$. Put $L=\liminf -n^{-1}\log q_n\in[0,+\infty]$,
with $-\log0=+\infty$. Every finite $a<L$ satisfies
$q_n\le e^{-na}$ eventually. Both terms then give a universal exponent
at least $a$. Let $a$ increase to $L$, or without bound if $L=+\infty$.
A zero value of $L$ is covered by nonnegativity. Finally, ML minimizes
error for each realized labelled codebook with uniform messages, so
$q_n\le\overline P_{e,n}(D_J)$ before and after codebook averaging.
This proves the reverse exponent inequality and completes the proof.
\end{proof}

\subsection{Gaussian likelihood-ratio comparison}\label{app:transfer}
\begin{proof}[Proof of Lemma~\ref{lem:gaussian-transfer}]
The sample space includes the codebook, transmitted label, received word,
and decoder randomness. The two experiments differ only in the received
conditional Gaussian mean. Including these common variables lets the likelihood-ratio bound apply
directly to the ensemble-average error event. The same argument then
handles a deterministic codebook as a special case.

We use $\EE|Z_j|^2=1$ for each standard proper complex Gaussian entry.
Whiten the noise. Conditional on the message and codebook, let
$h=f_1-f_2$ in whitened coordinates and let $Z\sim\CN(0,I)$ under
experiment 2. The likelihood ratio satisfies
\begin{align*}
 \log\mathcal L&=2\Re\langle Z,h\rangle-\norm h^2,\\
 \EE_2\mathcal L^\nu&=e^{\nu(\nu-1)\norm h^2}\le e^{\nu(\nu-1)\Delta^2}.
\end{align*}
Indeed $2\Re\langle Z,h\rangle$ is real Gaussian with variance
$2\norm h^2$. The moment bound remains valid after averaging over the
common message and random codebook; common independent tie randomness
can be included in that experiment. H\"older's inequality on its error
event gives~\eqref{eq:gaussian-holder} for $\nu>1$.
Exchanging the experiments gives the same bound in the other direction.

For changing ML rules, use experiment 2's ML decoder in experiment 1.
Optimality there and \eqref{eq:gaussian-holder} give
$e_{1,\mathrm{ML}}\le
e^{(\nu-1)\Delta^2}e_{2,\mathrm{ML}}^{1-1/\nu}$.
Exchange experiments for the reverse bound. This applies directly to the
joint random-codebook experiment.
\end{proof}

\subsection{Merging and normalization}\label{app:merging}
PEP puts different metrics on a common scale: the canonical score
$-\log p_\theta(x,y)$ measures how unlikely it is that a random word
defeats $x$ under metric $g_\theta$. The generalized minimum-error test
of~\cite[Sec.~IV]{elkayam} uses the merged score
\begin{equation}\label{main:eq:merge}
 J_{\rm merge}(x,y)
 =\sup_\theta[-\log p_\theta(x,y)]
 =-\log\inf_\theta p_\theta(x,y).
\end{equation}
It compares candidates through their smallest competitor probability
across the family. A sufficient normalization is
$\sup_y\EE_{X'}e^{J_{\rm merge}(X',y)}\le e^{o(n)}$.
Markov's inequality then bounds its PEP by
$e^{-J_{\rm merge}(x,y)+o(n)}\le e^{o(n)}p_\theta(x,y)$ for every
$\theta$, giving the required comparison.
For a finite family of subexponential size, uniform normalization of
the canonical scores gives this bound and yields the merged decoder
of~\cite[Sec.~IV-B, Theorem~2]{elkayam}.

In~\cite{elkayam}, we established the finite-family merging result
under a normal-prior assumption, which controls the expectation of the
inverse PEP. A simpler approach developed
in~\cite[Sec.~IV-A, Lemma~14]{elkayam-pep} caps the inverse PEP when
the PEP is very small. For any coding prior, any fixed output $y$,
and any metric $g$, the bound is
\[
 \EE_{X'\sim Q}\min\{p_g(X',y)^{-1},\alpha\}
 \le 1+\log\alpha,\qquad \alpha>1.
\]
Thus a cap of $e^{n^2}$ gives a subexponential moment bound without
requiring a normal prior. The newer merging proof uses this bound for the clipped score and
bounds the error caused by clipping.

Following the substitute-score approach of~\cite[Sec.~IV-C]{elkayam},
we use the statistic density to construct our score. We prove its moment bound directly.
Lemma~\ref{main:prop:criterion} compares it with the merged score below
a threshold and bounds the error above that threshold separately. The latter error decreases faster
than exponentially. As in the newer proof, this lets us handle very high
scores separately. Here, however, the threshold is used only in the proof;
the decoder uses the full score. The precise certificate is stated in Section~\ref{main:sec:framework}.

\section{Statistic Density Calculations}\label{sec:density}\label{app:density}
\subsection{Bases, reference volume and competitor boundary}\label{app:geometry}
Lemma~\ref{lem:statistic-domain} reconstructs every feasible statistic
and defines the reference measure. The reference volume is the one
induced by its inner product, on either the single total-trace space or
the space with one trace constraint per block.

Changing $W_i$ to another orthonormal basis changes $z_i$ by right
unitary multiplication, which preserves both $z_iz_i^*$ and volume.
The projector $Y_i^*(Y_iY_i^*)^{-1}Y_i$ uses an ordinary inverse
on the full-row-rank event. The residual score is measurable without
choosing a global row-space basis.

Write $a_*=\min_i a_i\asymp n/B$ and $a_{\rm tot}=\sum_i a_i\le n$.
The two ensembles have the same determinant powers, but different
normalizers, affine supports, and induced volumes. In the following,
$p,C,\mathcal A,\Omega,d$ refer to the selected ensemble. The reconstruction in Lemma~\ref{lem:statistic-domain}
applies to either trace constraint and realizes every feasible
statistic. Continuity and positive density in the interior give
the claimed support, including its closure. Each boundary component is
the zero set of the polynomial $\det(f_i-z_iz_i^*)$ on $\mathcal A$.
It is not the zero polynomial: take $z_i=0$ and positive scalar input
Grams with the prescribed traces (or any positive allocation of the
total trace). Thus the boundary has zero induced affine volume and zero
prior probability. Logarithms are evaluated on the relative interior.

This is the fixed-output competitor nullity used by
Lemma~\ref{lem:universal-pep}, including its zero-density convention.
The coordinate calculations in Appendix~\ref{app:spherical-density}
provide both densities at every prescribed positive energy.

\subsection{Gaussian projection and Wishart joint density}\label{app:gaussian-density}
\begin{proof}[Proof of Lemma~\ref{lem:gaussian-statistic}]
Complete $W$ to a unitary matrix $[W\ U]$ and write
\[
 [Z'\ V']=G[W\ U],\qquad V'=GU\in\CC^{t\times k},\quad k=m-s.
\]
Gaussian unitary invariance makes $Z'$ and $V'$ independent standard
complex Gaussian matrices. Therefore
\[
 F'=Z'Z'^*+V'V'^*,\qquad R'=V'V'^*.
\]
The standard complex Wishart density~\cite{goodman} gives
\begin{equation}\label{eq:wishart}
 p_{R'}(q)=C_{k,t}e^{-\tr q}\det(q)^{k-t},\qquad q\succ0,
\end{equation}
relative to the chosen Hermitian volume, for a positive constant
$C_{k,t}$. Since $Z'$ is independent of $V'$, it is independent of
$R'$. Multiplying by its Gaussian density gives a constant times
$e^{-\tr q-\norm z_F^2}\det(q)^{k-t}$. For fixed $z$, the translation
$f=q+zz^*$ preserves volume. Substitution gives
$\tr q+\norm z_F^2=\tr f$ and proves~\eqref{eq:single-gaussian}.
\end{proof}

For independent blocks the densities multiply, giving
\begin{equation}\label{eq:gaussian-product}
 p_G(f,z)=C_Ge^{-\sum_i\tr f_i}\prod_i\det R_i^{a_i},
\end{equation}
where $C_G>0$ depends only on the block dimensions.

\subsection{Global and per-block radial normalization}\label{app:spherical-density}
\begin{proof}[Proof of Lemma~\ref{lem:spherical-density}]
Write the concatenated Gaussian input as $G=\sqrt S\,U$, where
$S=\norm G_F^2$. The direction $U$ is uniform on the unit sphere and
independent of $S$. Thus $X=\sqrt{E/S}\,G=\sqrt E\,U$ has the required
spherical law for every prescribed $E>0$. Its statistics satisfy
\begin{equation}\label{eq:radialscaling}
 F_i=(E/S)F_i',\qquad Z_i=\sqrt{E/S}\,Z_i'.
\end{equation}

Change coordinates from $(f',z')$ to energy $s>0$ and normalized
statistic $u=(f,z)$ on the energy-$E$ trace space. The inverse is
$f_i'=(s/E)f_i$, $z_i'=\sqrt{s/E}\,z_i$. Choose one Hermitian
coordinate along the total trace and all others tangent to its level
sets. In these coordinates the derivative is block triangular.
The derivative with respect to energy can depend on $u$, but this
block lies below the diagonal and does not affect the determinant.
At fixed $s$, every Hermitian tangent direction is scaled by $s/E$
and every real projection direction by $\sqrt{s/E}$. Therefore the
Jacobian depends on $s$ and the dimensions, not on $u$.

The Gaussian exponential is $e^{-s}$ and
$R_i'=(s/E)R_i$. Substituting into~\eqref{eq:gaussian-product}, the
joint energy/statistic density has the form
\[
 h_E(s)\prod_i\det R_i(u)^{a_i},\qquad s>0,\quad u\in\Omega_{\rm w},
\]
where $h_E(s)>0$ is independent of $u$. Integrating over $s$ gives
\eqref{eq:whole-density}, with a positive constant $C_{\rm w}$ making
the density integrate to one. This deterministic normalization map
works for each $E>0$; no conditioning on a zero-probability event is used.

For independent block normalization, apply the same calculation to each
$X_i=\sqrt{E_i/S_i}\,G_i$. The normalized blocks remain independent,
so their densities multiply and give~\eqref{eq:block-density} for every
prescribed positive energy vector. Both constants depend only on
dimensions and energies. Gaussian unitary invariance removes any
dependence on output magnitudes or the choice of row-space bases.
\end{proof}

\section{Residual and Local-Density Estimates}\label{app:local}
Use the selected ensemble's notation from Section~\ref{main:sec:universality}.
Put $a_*=\min_i a_i\asymp n/B$ and $a_{\rm tot}=\sum_i a_i\le n$.
All estimates hold uniformly over full-row-rank outputs and auxiliary channels.

\subsection{The residual eigenvalue floor}\label{app:residual-floor}
Use the selected ensemble's normalized density. The reference-ball
argument in Section~\ref{main:sec:universality} gives
$\log C\le O(n)$ in~\eqref{eq:normalizer-bound}.

\begin{proof}[Proof of Lemma~\ref{lem:residual-floor}]
We now convert the score bound into a deterministic residual floor.
Total energy bounds the other eigenvalues from above, limiting their
contribution to the log-determinants. Isolating the smallest eigenvalue
then bounds how far that eigenvalue can decrease at a fixed score.

Use the proof cutoff~\eqref{eq:cutoff}.
For $u\in\mathcal G_n$, every eigenvalue of every residual is at most
$E$, because $0\preceq R_i\preceq f_i$ and $\tr f_i\le E$.
Let $\rho(u)$ be the smallest of these eigenvalues, attained in block $j$.
Singling out that eigenvalue in the weighted sum gives
\[
 \sum_i a_i\log\det R_i
 \le a_j\log\rho(u)+(ta_{\rm tot}-a_j)\log E.
\]
Using the definition of $J$, we obtain the central inequality
\begin{equation}\label{eq:direct-eigen}
 a_j\log\frac1{\rho(u)}
 \le J(u)+\log C+(ta_{\rm tot}-a_j)\log E.
\end{equation}
The right side is bounded above by $O(n\log n)$ on $\mathcal G_n$,
using \eqref{eq:normalizer-bound}, $ta_{\rm tot}\le tn$, and $E=nP$.
Since $a_j\ge a_*\asymp n/B$,
\begin{equation}\label{eq:eigen-floor}
 \rho(u)\ge\exp[-C_{\rho}B\log n]
 \quad\text{uniformly on }\mathcal G_n.
\end{equation}
Here $C_{\rho}$ is a sufficiently large fixed constant.

\end{proof}

\section{Energy Moments and Movement}\label{app:energy}
\subsection{Gamma moments and the displacement bound}\label{app:movement}
\begin{proof}[Proof of Lemma~\ref{lem:energy-movement}]
The independent Gaussian squared radii have gamma shapes $tm_i$.
With $\Lambda_i=S_i/\sum_j S_j$, the energy proportions have the
Dirichlet distribution with parameters
$tm_1,\ldots,tm_B$. This is the standard normalization of independent
gamma variables~\cite[Ch.~XI, Thm.~4.1]{devroye}. Its density is
\begin{equation}\label{eq:dirichlet-new}
 \frac{\Gamma(tn)}{\prod_i\Gamma(tm_i)}
 \prod_i u_i^{tm_i-1},\quad u_i>0,\quad\sum_i u_i=1,
\end{equation}
relative to $du_1\cdots du_{B-1}$. We include the change of variables
and the moment calculation for completeness. To verify the density, use
$S_i=v u_i$ with $v=\sum_i S_i$. Its Jacobian is $v^{B-1}$;
the joint gamma density factors into a shape-$tn$ gamma density for $v$
and \eqref{eq:dirichlet-new}. Integrating that density after increasing
one exponent by one or two gives the moments in~\eqref{eq:energy-moments}.
Set
\[
 \mathfrak d(\Lambda)=E^{-1}\norm{\Phi_n(X)-X}_F^2
 =\sum_i(\sqrt{\Lambda_i}-\sqrt{e_i})^2.
\]
For $u\ge0$ and $e>0$, rationalizing the difference of square roots gives
\[
 (\sqrt u-\sqrt e)^2
 =\frac{(u-e)^2}{(\sqrt u+\sqrt e)^2}\le\frac{(u-e)^2}{e}.
\]
The energy fractions have exactly their natural values in expectation.
Thus the squared deviations on the right average to variances, with no
bias term. Using this inequality and
\eqref{eq:energy-moments},
\begin{equation}\label{eq:aggregate-bound}
 \EE[\mathfrak{d}(\Lambda)]\le\sum_i\frac{\operatorname{Var}(\Lambda_i)}{e_i}
 =\frac{B-1}{tn+1}=: \varepsilon_n.
\end{equation}
For $B>1$, retain the energy-only event
$\mathcal E_n=\{\mathfrak d(\Lambda)\le2\varepsilon_n\}$. Markov's inequality gives
$\PP\{\mathfrak d(\Lambda)>2\varepsilon_n\}\le1/2$, so
\begin{equation}\label{eq:energy-event}
 \pi_n:=Q_1(\mathcal E_n)\ge\tfrac12.
\end{equation}
On this event,
\begin{equation}\label{eq:input-movement}
 \norm{\Phi_n(X)-X}_F^2\le 2nP\varepsilon_n=o(n).
\end{equation}
Only $B=o(n)$ is needed for this step; no simultaneous relative-energy
concentration for every block is required. When $B=1$ the map is the
identity.

Finally, the frequency-domain whitened operator has norm at most
$G_{\theta_n}$. Multiplying the input displacement bound by its square
proves~\eqref{eq:mean-movement}. The gain condition makes this $o(n)$.
\end{proof}

\section{Additional Calculations and Guarantees}\label{app:interpretation}
\subsection{Finite-family codebook guarantees}\label{app:finite-family}
For a finite channel family $\Theta_n$ let $\mu_{j,n}$ be the
ensemble-average error of $D_J$. Markov's inequality and the union bound
give
\[
 \PP_{\mathcal C}\{\exists j:\ P_e(\mathcal C,j)
       >e^{n\eta_n}\mu_{j,n}\}\le |\Theta_n|e^{-n\eta_n}.
\]
If $\log |\Theta_n|=o(n)$, take $n\eta_n=\log |\Theta_n|+\sqrt n$. Then
$\eta_n\to0$ and a deterministic codebook meets all these ensemble
benchmarks up to a subexponential factor. Extending this conclusion from
a finite net to a continuum requires approximation and uniformity on the
error-exponent scale. The fixed-channel ensemble theorem and this
finite-family extraction give the corresponding two levels of guarantee,
as discussed in
\cite{feder-lapidoth,lapidoth-narayan}.

\end{document}